\documentclass[twocolumn,aps,amssymb,prl,showpacs]{revtex4}
\usepackage{graphicx}
\usepackage{amsmath}

\begin{document}

\title{Direct observation of Josephson capacitance}

\author{M. A. Sillanp\"a\"a$^1$, T.
Lehtinen$^1$, A. Paila$^1$, Yu. Makhlin$^{1,2}$, L. Roschier$^1$,
and P. J. Hakonen$^1$} \affiliation{ $^1$Low Temperature
Laboratory, Helsinki University of Technology,
FIN-02015 HUT, Finland \\
$^2$Landau Institute for Theoretical Physics, 119334 Moscow,
Russia}


\begin{abstract}
The effective capacitance has been measured in the split Cooper
pair box (CPB) over its phase-gate bias plane. Our low-frequency
reactive measurement scheme allows to probe purely the capacitive
susceptibility due to the CPB band structure. The data are
quantitatively explained using parameters determined independently
by spectroscopic means. In addition, we show in practice that the
method offers an efficient way to do non-demolition readout of the
CPB quantum state.
\end{abstract}

 \bigskip

\maketitle

Energy can be stored into Josephson junctions (JJ) according to $E
= -E_J \cos (\varphi)$, where $\varphi$ is the phase difference
across the junction, and the Josephson energy $E_J$ is related to
the junction critical current $I_C$ through $I_C = 2e E_J /\hbar$.
By using the Josephson equations, this energy storing property
translates into the well-known fact that a single classical JJ
behaves as a parametric inductance $L_{J} = \hbar /(2 e I_C)$ for
small values of $\varphi$.

Since the early 80's, it has become understood that $\varphi$
itself can behave as a quantum-mechanical degree of freedom
\cite{Tinkham}. In mesoscopic JJs, this is typically associated
with the competition between the Josephson and Coulomb effects at
a very low temperature. These fundamental phenomena take place if
charge on the junction is localized by a large resistance $R > R_Q
= h/(4e^2)$ \cite{SZ}, as well as in the Cooper-pair box (CPB), or
the single-Cooper-pair transistor (SCPT), whose quantum coherence
is often considered macroscopic~\cite{Nakamura}.

In the first theoretical landmark papers \cite{widom84,Likharev85}
on quantum properties of $\varphi$ it was already noticed that due
to localization of charge $Q$, the energy of a the JJ system is
similar to that of a non-linear capacitance. In spite of the
importance of the phenomenon especially in CPB or SCPT in the
promising field of superconducting qubits
\cite{AverinBruder03,MSS}, direct experimental verification of the
Josephson capacitance has been lacking, likely due to challenges
posed by measuring small reactances, or by the extreme sensitivity
to noise.

In this Letter, we present the first such direct experiment
\cite{Duty05}, where we determine the Josephson capacitance in the
Cooper pair box. Related experiments have recently been performed
by Wallraff \emph{et~al.}~\cite{Wallraff04}, but in their case the
key role is played by the transitions between levels of a coupled
system where the band gap between the ground state and first
excited state of the CPB, $E_1 - E_0 = \Delta E$, is nearly at
resonance with an oscillator of angular frequency $\omega_0$.
Thus, detuning fully dominates over the Josephson capacitance
which can be clearly observed in our experiments where we study
directly the reactive response of the lowest band $E_0$ at a
frequency $\omega_0 \ll \Delta E/\hbar$. We determine the
experimental parameters independently using spectroscopy, and
demonstrate a simple way to perform a non-destructive measurement
of the CPB state using purely the CPB Josephson capacitance.

\begin{figure}

    \includegraphics[width=8cm]{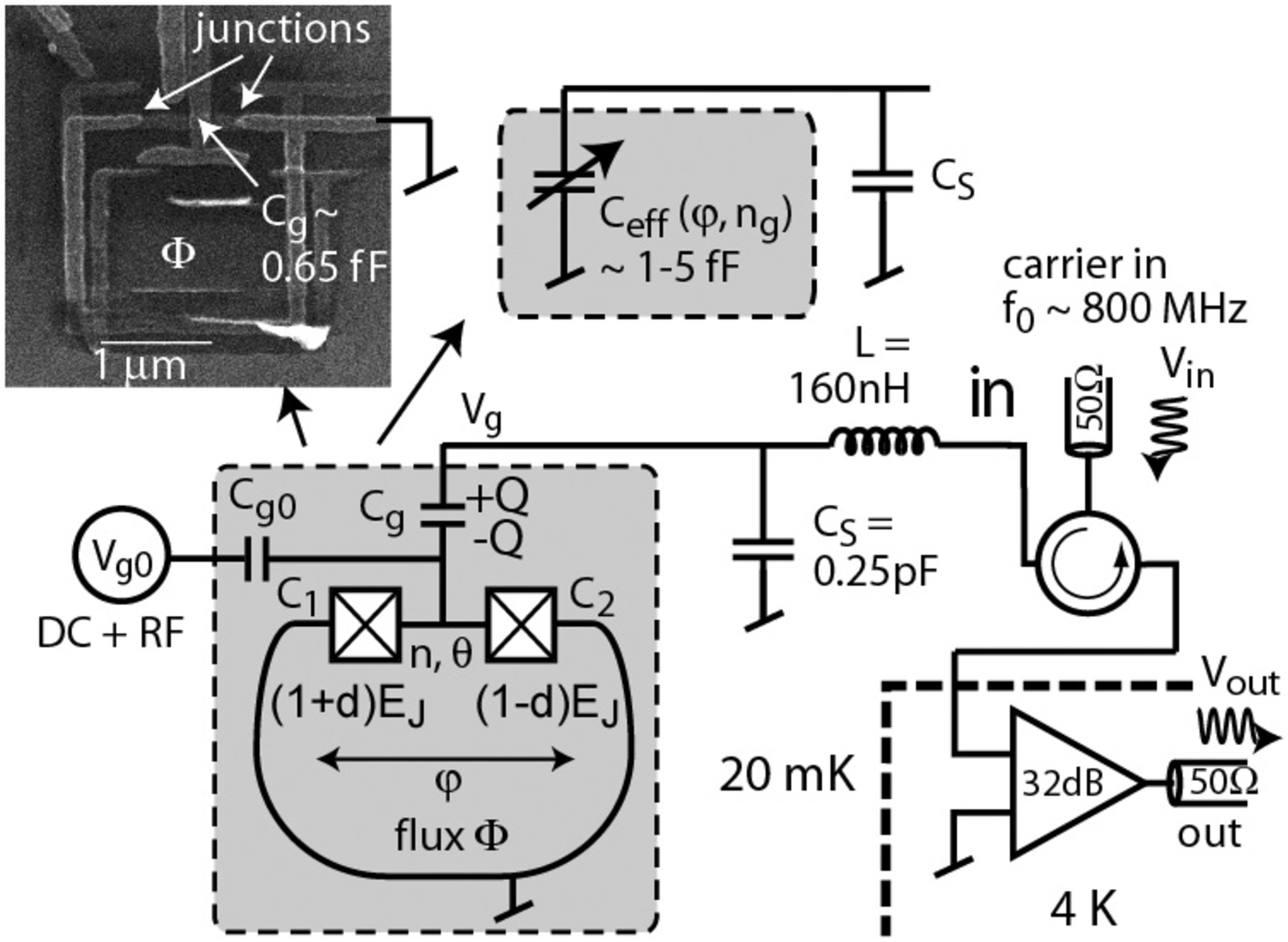}

    \caption{Schematic view of the experiment. The resonant
    frequency of the $LC$ circuit (made using lumped elements)
    is tuned by the effective capacitance $C_\mathrm{eff}$ of the Cooper pair box shown
    in the SEM image. For details, see text. }\label{Scheme}

    \end{figure}

An SCPT (Fig.~\ref{Scheme}) consists of a mesoscopic island (total
capacitance $C = C_1 + C_2 + C_g$), two JJs, and of a nearby gate
electrode used to polarize the island with the (reduced) gate
charge $n_g = C_g V_g / e$. The island has the charging energy
$E_C = e^2/(2C)$, and the junctions have the generally unequal
Josephson energies $E_J(1 \pm d)$, where the asymmetry is given by
$d$. The SCPT Hamiltonian is then $E_C (\hat{\mathrm{n}}-n_g)^2 -
2E_{J} \cos \left(\varphi/2 \right) \cos (\hat{\theta}) + 2d
E_{J}\sin \left(\varphi/2 \right) \sin (\hat{\theta}) - C_g
V_g^2/2$. Here, the number $\hat n$ of extra electron charges on
the island is conjugate to $\hat\theta/2$, where $\hat\theta$ is
the superconducting phase on the island \cite{HamCom}. The SCPT is
thus equivalent to a CPB (single JJ and a capacitance in series
with a gate voltage source) but with a Josephson energy tunable by
$\varphi = 2 \pi \Phi / \Phi_0$, where $\Phi_0 = h/(2e)$ is the
quantum of magnetic flux.

If $d = 0$ and $E_J / E_C \ll 1$ the ground and excited state
energies are ($n_g = 0 ... 2$): $E_{0,1} = E_C (n_g^2 - 2 n_g + 2)
\mp \sqrt{\left(E_J \cos(\varphi/2) \right)^2 + \left(2 E_C
(1-n_g) \right)^2} - C_g V_g^2/2$, with a large gap to higher
levels. For a general $E_J/E_C$, we compute the bands numerically
in the charge state basis.

The effective "Josephson" capacitance of the CPB can be related to
the curvature of band $k$, similar to the effective mass of an
electron in a crystal:
\begin{equation}
C_{\rm eff}^k = - \frac{\partial^2 E_k(\varphi, n_g)}{\partial
V_g^2} = - \frac{C_g^2}{e^2} \frac{\partial^2 E_k (\varphi,
n_g)}{\partial n_g^2}\,. \label{Ceff_defin}
\end{equation}
Usually, the system effective capacitance is obtained from a
Lagrangian or Hamiltonian as $\partial^2 {\cal L}/\partial V_g^2 =
(\partial^2 H / \partial Q^2)^{-1}$, without the minus sign. In
Eq.~(\ref{Ceff_defin}), however, $E_k$'s are, more precisely, the
eigenvalues of the {\it Routhian} ${\cal H} = \dot\theta
\partial_{\dot\theta}{\cal L} -{\cal L}$~\cite{LandauLifschitzI},
which serves as a Hamiltonian for the $n, \theta$ degree of
freedom but as {\it minus} Lagrangian for the phase $\alpha\equiv
\frac{e}{\hbar} \int V_g dt$ and $V_g \propto \dot\alpha$, thus
leading to Eq.~(\ref{Ceff_defin}).

Using the analytic formulas for $E_{0,1}$ in the limit $E_J / E_C
\ll 1$ we get
\begin{equation}\label{CeffAnalyt}
\begin{split}
    & C_\mathrm{eff}^{(0,1)} =  C_g - \frac{2 C_g^2 E_C}{e^2} \times \\
    & \times \left(1 \mp \frac{E_C E_J^2 (1+\cos \varphi)}{\left[4 E_C^2 (n_g-1)^2 +
\frac{1}{2} E_J^2 (1+\cos \varphi) \right]^{3/2}} \right),
   \end{split}
\end{equation}
which reduces to the classical geometric capacitance $(1/C_g
+1/(C_1 + C_2))^{-1}$ in the limit of vanishingly small $E_J$,
except where Cooper-pair tunneling is degenerate \cite{Stafford}.
Numerically evaluated graphs of $C_\mathrm{eff}^{(0,1)}$ for a
general $E_J/E_C$ can be found in Ref.~\cite{MIKAthesis}.

Our experimental scheme is illustrated in Fig.~\ref{Scheme}. We
perform low-dissipation microwave reflection measurements
\cite{lset,Leif05} on a series $LC$ resonator in which the box
effective capacitance, Eq.~(\ref{Ceff_defin}), is a part of the
total capacitance $C_S+C_{\mathrm{eff}}^{k}$. The resonator is
formed by a surface mount inductor of $L = 160$ nH. With a stray
capacitance of $C_S = 250$ fF due to the fairly big lumped
resonator, the resonant frequency is $f_0 = 800$ MHz and the
quality factor $Q \simeq 16$ is limited by the external $Z_0 = 50
\, \Omega$. When $C_{\mathrm{eff}}^{k}$ varies, the phase $\Theta$
of the reflected signal $V_{\mathrm{out}}=\Gamma V_{\mathrm{in}}$
changes, which is measured by the reflection coefficient $\Gamma =
(Z-Z_0)/(Z+Z_0)=\Gamma_0 e^{i\Theta}$. Here, $Z$ is the resonator
impedance seen at the point labelled "in" in Fig.~\ref{Scheme}. In
all the measurements, the probing signal $V_{\mathrm{in}}$ was
continuously applied.

Since we are rather far from matching conditions, the reflection
magnitude $\Gamma_0$ remains always close to one. The variation in
$\Theta$ due to modulation in $C_{\mathrm{eff}}^{k}$ is up to
40$^{\circ}$ in our measurements, corresponding to a shift of
resonance frequency $\Delta f_0 \simeq 6$ MHz. In addition to band
pass filtering, we used two circulators at 20 mK.

As seen in Eq.~(\ref{CeffAnalyt}), the modulation depth of
$C_{\mathrm{eff}}^{k}$ is sensitive to $C_g$. Therefore, in order
to faithfully demonstrate the Josephson capacitance in spite of
the stray capacitance, we used a large $C_g > 0.5$ fF. It was made
using an Al-AlO$_x$-Al overlay structure (see the image in
Fig.~\ref{Scheme}), with a prolonged oxidization in 0.1 bar of
O$_2$ for 15 min. Otherwise, our CPB circuits have been prepared
using rather standard e-beam lithography. The tunnel junctions
having both an area of 60 nm $\times$ 30 nm correspond to an
average capacitance of $\sim$ 0.17 fF each. The overlay gate has
$C_g \simeq 0.7$ fF for an area of 180 nm $\times$ 120 nm.

The main benefit of our method comes from the fact that we work at
a resonator (angular) frequency $\omega_0$ much lower than the CPB
level spacing $\Delta E$. In Ref.~\cite{Wallraff04} it is shown
that $\omega_0$ depends on the resonator - CPB (qubit) interaction
because of two contributions. The frequency change is $\Delta
\omega_0 = g^2 / \delta$, where the detuning $\delta = \Delta E  -
\hbar \omega_0$, and the coupling coefficient $g$ contains the
curvature of energy bands. In general, both $\delta$ and the
curvature depend on the $(n_g, \, \varphi)$ point. Now, in our
case everywhere $\Delta E \gg \hbar \omega_0$, $\delta \simeq
\Delta E$, and hence $\Delta \omega_0 = g^2 / \Delta E =
C_{\mathrm{eff}} \omega_0 /(2 C_S)$ has a contribution by only the
second derivative, not by the detuning. Therefore, we can resolve
the reactive response due to purely the bands of CPB, which has
not been possible in previous experiments.

When doing microwave spectroscopy, we have to consider also the
other side of the coin: $\Delta E$ increases due to interaction
with the resonator by \cite{Wallraff05} $\varepsilon = \hbar
\left(2 N g^2/\Delta E + g^2/\Delta E \right)$, where $N$ is the
number of quanta in the resonator. When driven by a gate amplitude
$V_g$, the resonator energy is $E_R = V_g^2 C_S/2$. At a high
excitation amplitude $n_g \simeq 1/2$ we would have $V_g \simeq
e/(2 C_g )$ and hence $N = E_R /(\hbar \omega_0) = e^2C/(8C_g^2
\hbar \omega_0) \sim 4 \times 10^3$ which would yield $\varepsilon
\sim \Delta E$. The data shown in this paper are, however,
measured at a very low excitation of $n_g \sim 0.05$ which
corresponds to $N \sim 40$ and $\varepsilon \sim 200$ MHz which is
an insignificant contribution to $\Delta E$.

\begin{figure}[h]
    \includegraphics[width=7.0cm]{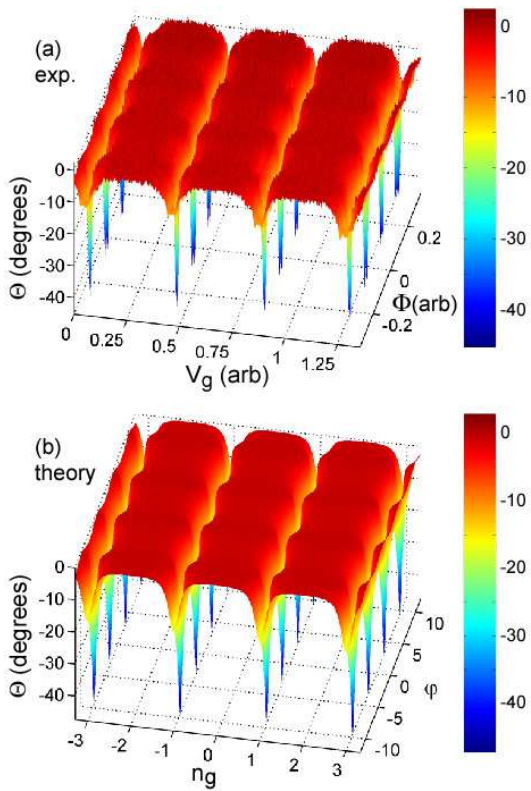}

    \caption{(color online) (a) Phase shift $\Theta$ measured at a probing frequency 803 MHz $\sim f_0$,
    and (b) $\Theta$ calculated
    using Eq.~(\ref{Ceff_defin}) with the ground band $E_0(\varphi, n_g)$ evaluated numerically
    with parameters of Table
    \ref{tb:samples}.} \label{ResultPhase}
    \end{figure}

Fig.~\ref{ResultPhase} (a) displays the measured phase shift
$\Theta$ as a function of the two external control knobs (in the
following, $n_g$ should be understood as being due to the control
gate, $n_g = C_{g0} V_{g0}/e$). The results show full $2e$
periodicity as a function of $n_g$, checked by increasing
temperature above the $2e - e$ crossover at $\sim 300$ mK, and a
$\Phi_0$ period with respect to $\Phi$. The data was measured
without any microwave excitation, and hence we expect to see
effects due to the ground band $C_{\mathrm{eff}}^{0}$. The
corresponding theoretical picture, obtained using
Eq.~(\ref{Ceff_defin}) and straightforward circuit formulas for
$\Gamma$, is given in Fig.~\ref{ResultPhase} (b).

\begin{figure}[h]

    \includegraphics[width=8.5cm]{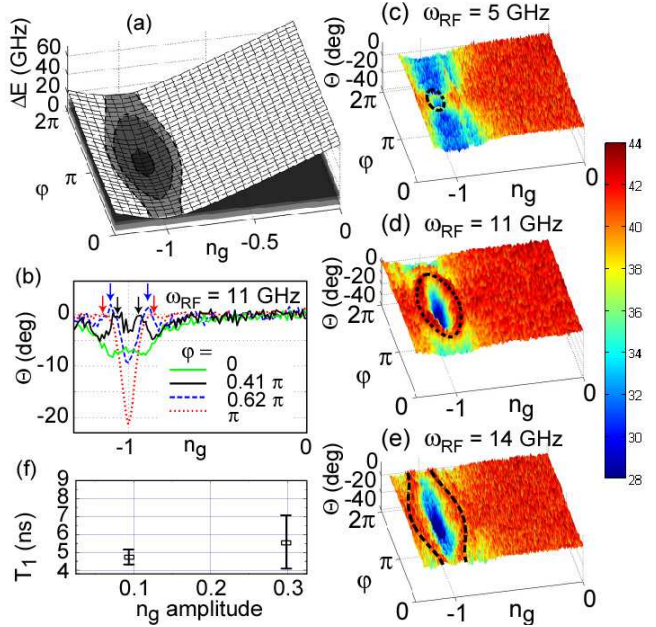}

    \caption{(color online) (a) Illustration of the microwave spectroscopy used to map the SCPT
    band gap $\Delta E = E_1 - E_0$. The three horizontal planes
    which intersect the band gap correspond, from bottom to top,
    to the microwave energy $\hbar \omega_{RF}$ used in (c)...(e), respectively.
    Whenever $\Delta E = \hbar \omega_{RF}$ (dashed lines), the system experiences
    resonant absorption; (b) peaks of resonant absorption (arrows) in
    the measured phase shift at $\omega_{RF} = 11$ GHz;
    (c) - (e) spectroscopy data represented as surfaces in
    the $\varphi$, $n_{g}$ plane. The resonance conditions shown in (a) are
    plotted on top of the data; (f) $T_1$ as a function of measurement strength at
    $\varphi = 0$, $n_g \sim 1$.}\label{fig:spectroscopy}
    \end{figure}

As a vital step to get convinced of the measured capacitance
modulation versus the calculation, we carried out a detailed
determination of the sample parameters independently of the
capacitance modulation by using microwave spectroscopy
(Fig.~\ref{fig:spectroscopy}). To the weakly coupled control gate
$C_{g0}$ of the SCPT, we applied continuous-wave microwaves while
slowly sweeping the CPB band gap $\Delta E$ with $\varphi$ and
$n_{g}$. Whenever the microwave energy matches the band gap, that
is, $\hbar \omega_{\mathrm{RF}} = \Delta E$, the CPB becomes
resonantly excited. Since typically the Josephson corrections to
the geometric capacitance are opposite in sign for the bands 0 and
1 (see Eq.~(\ref{CeffAnalyt})), band 1 would contribute an
opposite phase shift signal. At resonance, we would then expect to
see mixture of $C_{\mathrm{eff}}^{0}$ and $C_{\mathrm{eff}}^{1}$,
weighted by the state occupancies which depend on the microwave
amplitude. We calculate that a high enough amplitude sufficient to
saturate the populations into a $50-50$ \% mixture, would yield a
$\sim 3^{\circ}$ resonance absorption peak in the measured
$\Theta$. The expectation is confirmed in
Fig.~\ref{fig:spectroscopy} (b), where the resonance peaks are
displayed at a few values of $\varphi$ (when $\varphi = 0$,
microwave energy does not exceed the band gap, and for $\varphi =
\pi$ the peak height is lower due to a smaller matrix element).

While slowly sweeping $\varphi$ and $n_{g}$, the resonance
conditions correspond to contours (see Fig.~\ref{fig:spectroscopy}
(a)), which appear as annular ridges in the experimental data of
graphs \ref{fig:spectroscopy} (c)-(e) around the minimum $\Delta
E$ at $(n_g = -1, \, \varphi = \pi)$. Since the band gap is
sensitive to $E_J$ as well as to the $E_J/E_C$ ratio, the
resonance contours allow for an accurate determination of these
parameters (Table~\ref{tb:samples}). For example, at $(n_g = -1,
\, \varphi = 0)$, the band gap is $2 E_J = 12.5$ GHz, whereas at
$(n_g = -1, \, \varphi = \pi)$ $\Delta E$ has the absolute minimum
$2 d E_J \simeq 3$ GHz which was barely exceeded by the microwave
energy in Fig.~\ref{fig:spectroscopy} (c).

Based on the surface area $\sim 0.022 \, (\mu$m$)^2$ of the
overlay gate, we estimate $C_g \sim 0.5 - 1$ fF. The exact value
was obtained by fitting to the modulation depth of
$C_{\mathrm{eff}}^{0}$ (see Eq.~(\ref{CeffAnalyt})), yielding $C_g
= 0.65$ fF, corresponding to a specific capacitance very
reasonable to a thick oxide $\sim 30$ fF/$\mu$m$^2$.

\begin{table}[h]
\begin{tabular}{|c|c|c|c|c|c|c|}
   \hline  $E_J$ (K) & $E_C = \frac{e^2}{2C}$ (K) & $E_J/E_C$ & $R_{T}(\mathrm{k} \Omega )$ & $C$ (fF) & $d$ & $C_g$ (fF)\\
   \hline 0.30 & 0.83  & 0.36 & 55 & 1.1 & 0.22 & 0.65 \\
      \hline
  \end{tabular}
  \caption{Sample parameters determined by microwave spectroscopy.
  $R_T$ is the series resistance of the two SCPT tunnel junctions
  (other parameters are defined in text and in Fig.~\ref{Scheme}).} \label{tb:samples}
\end{table}

\begin{figure}[h]

    \includegraphics[width=8.0cm]{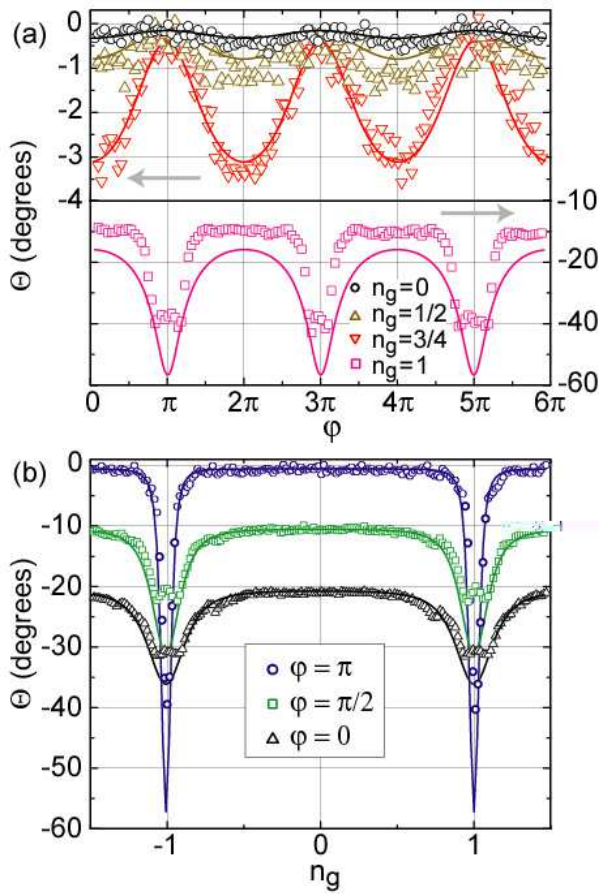}

    \caption{(color online) Measured microwave phase shift $\Theta$, (a) vs. phase $\varphi$
    (note different scales in the two panels) and
    (b) vs. gate charge $n_g$ (curves for $\varphi = 0$ and $\pi/2$ have been shifted vertically for
    clarity by 20$^{\circ}$ and 10$^{\circ}$, respectively). Solid lines are numerical calculations using
    Eq.~(\ref{Ceff_defin}) and sample parameters in Table \ref{tb:samples}.} \label{Modulation}
    \end{figure}

Fig.~\ref{Modulation} illustrates the bare gate and flux
modulations without microwave excitation in more detail, and shows
the corresponding numerical calculations using the ground band. As
expected, $C_{\mathrm{eff}}$ reduces to the geometric capacitance
when Cooper-pair tunneling is blocked either by tuning the
Josephson energy effectively to zero when $\varphi$ is an odd
multiple of $\pi$, or by gate voltage. At the Coulomb resonance
$n_g = \pm 1$, however, the Josephson capacitance is significant.
In the special point $n_g = \pm 1$, $\varphi = \pm \pi$, the most
pronounced effect is observed, now due to strong Cooper-pair
fluctuations. The agreement between theory and experiment is good
in Fig.~\ref{Modulation} except around $n_g = \pm 1$ which we
assign to intermittent poisoning by energetic quasiparticles
\cite{aumentado}. An estimate using $C_{\mathrm{eff}}$ from
Eq.~(\ref{CeffAnalyt}), $\Theta = -2 C_{\mathrm{eff}}
\sqrt{L}/(C_S^{3/2} Z_0)$, falls to within 15 \% of the numerical
results except around integer $n_g$.

Reactive measurements, either inductive or capacitive, can be
employed for non-demolition readout for qubits \cite{QND} which
means that $0 \leftrightarrow 1$ relaxation caused by the
measurement is insignificant. An important advantage of our scheme
is that since the probing gate swing has a frequency $f_0 \ll
\Delta E / \hbar$, its contribution to spectral density at the
qubit level spacing frequency is negligible. We measured the
relaxation time $T_1$ using the technique of pulsed microwave
excitation with variable repetition time $T_R = 1 - 200$ ns, while
keeping the measurement signal always on, as in
Ref.~\cite{Duty04}. The $T_1$ times were limited to about 7 ns by
parasitic reactances in the somewhat uncontrolled high-frequency
environment, causing noise from $Z_0$ to couple strongly due to a
large coupling $\kappa = C_g / C \sim 1$. The result for $T_1$,
however, did not depend on the measurement strength
(Fig.~\ref{fig:spectroscopy} (f)), which supports the
non-demolition character of this scheme. By fabricating the
resonator on-chip it is straightforward to gain a full control of
environment. Then, the impedance seen from the qubit gate
$\mathrm{Re}(Z_g(\omega = \Delta E/\hbar)) \simeq 0.1$ m$\Omega$,
and a worst-case estimate yields $T_1 \approx \hbar R_K / [4 \pi
\kappa^2 \mathrm{Re}(Z_g(\Delta E/\hbar)) \Delta E] \gg 1 \mu$s.
For a dephasing time $T_2$ averaged over $n_g$, we measured $\sim
0.5$ ns using Landau-Zener interferometry \cite{feigelman,Mika05}.
This $T_2$ time is on the same order as the spectroscopy line
widths in Fig.~\ref{fig:spectroscopy}.

In conclusion, using the phase of strongly reflected microwave
signals, we have experimentally verified the Josephson capacitance
in a mesoscopic Josephson junction, \textit{i.e.}, the quantity
dual to the Josephson Inductance. Good agreement is achieved with
the theory on the Josephson capacitance. Implications for
non-destructive readout of quantum state of Cooper-pair box using
the capacitive susceptibility are investigated.

We thank T. Heikkil\"a, F. Hekking, G. Johansson, M. Paalanen, and
R. Schoelkopf for comments and useful criticism. This work was
supported by the Academy of Finland and by the Vaisala Foundation
of the Finnish Academy of Science and Letters.




\begin{thebibliography}{99}
\bibitem{HamCom} The last term in the Hamiltonian typically does
not influence the SCPT's dynamics and is typically omitted; it is
important to keep it here since $V_g$ becomes a dynamical
variable.
\end{thebibliography}
\end{document}